\RequirePackage{fix-cm}
\documentclass{svjour3}                     
\smartqed  
\usepackage[caption=false]{subfig}
\usepackage{floatrow}
\usepackage{graphicx}
\usepackage{natbib,amsfonts,amssymb}
\usepackage[section]{placeins}
\usepackage{times} 
\usepackage[utf8]{inputenc}
\usepackage{color}

\begin{document}

\title{A numerical investigation of coorbital stability and libration in three dimensions  
}


\author{M. H. M. Morais         \and
        F. Namouni 
}


\institute{M. H. M. Morais \at
              Instituto de Geoci{\^e}ncias e Ci{\^e}ncias Exatas, Universidade Estadual Paulista, Av. 24-A, 1515 13506-900 Rio Claro, SP, Brazil \\
                    \email{helena.morais@rc.unesp.br}           
           \and
           F. Namouni \at
              Universit\'e C\^ote d'Azur, CNRS, Observatoire de la C{\^o}te d'Azur, CS 34229, 06304 Nice, France
}

\date{Received: date / Accepted: date}

\maketitle

\begin{abstract}
Motivated by the dynamics of resonance capture, we study numerically the coorbital resonance for  inclination $0\leq I\leq180^\circ$ in the circular restricted three-body problem. We examine the similarities and differences between planar and three dimensional  coorbital resonance capture and seek their origin in the stability of coorbital motion at arbitrary inclination. After we present stability maps of the planar prograde and retrograde coorbital resonances, we characterize the new coorbital modes in three dimensions. We see that  retrograde mode I (R1) and mode II (R2)  persist as we change the relative inclination, while retrograde mode III (R3) seems to exist only in the planar problem.
A new coorbital mode (R4) appears in 3D which is a retrograde analogue to an horseshoe-orbit.  
The Kozai-Lidov resonance is active for retrograde orbits as well as prograde orbits and plays a key role in coorbital resonance capture.
Stable coorbital modes exist at all inclinations, including retrograde and polar obits. This result confirms the robustness the coorbital resonance at large inclination and encourages the search for retrograde coorbital companions of the solar system's planets.
\keywords{Co-orbital resonance \and Resonance \and Three-body problem }
\end{abstract}

\section{Introduction}
The recent discovery of the Centaurs and Damocloids 2006 BZ8, 2008 SO218, 2009 QY6 and 1999 LE31 in retrograde resonance with Jupiter and Saturn \citep{Morais&Namouni2013MNRASL} has shown that retrograde resonances although weaker than their prograde counterparts \citep{Morais&Namouni2013CMDA} (hereafter Paper I) are capable of temporarily trapping minor bodies for a few thousand years at a time. A subsequent statistical study of resonance capture at arbitrary inclination \citep{Namouni&Morais2015MNRAS} (hereafter Paper II) reported the unexpected result that although weaker than prograde resonances, retrograde resonances have a higher capture likelihood. Since resonance provides a protection mechanism against disruptive close encounters with the planets, the current Centaurs and Damocloids in retrograde resonance could be among the dynamically oldest minor bodies in the outer solar system. It is therefore of interest to study closely the dynamics of such resonances in terms of both capture and stability. In this regard, the strongest resonance for retrograde motion is the coorbital (1:1) resonance. Moreover, it was shown in Paper II how inwardly drifting outer particles  that encounter Jupiter's web of mean motion resonances with a small relative inclination are captured with great likelihood in the coorbital resonance if their orbits are retrograde, and in the 1:2 resonance if their orbits are prograde. For this reason, in this paper we investigate further the workings of capture and stability in the retrograde coorbital resonance.  

In section  2 we start  by reviewing our work presented in Paper I on the coorbital resonant angles at arbitrary inclination and the possible coorbital modes in 2D. We progress by examining the similarities and differences between capture in the planar and three dimensional configurations (section  3) as dynamical models of coorbital motion often rely on the planar problem to access the basics of resonance capture.
We show that the two configurations are fundamentally different owing to the presence  of the Kozai resonance in 3D that conditions the three-stage capture process reported in Paper II. To further understand how specific librations are possible in the planar and three-dimensional configurations, we produce (2D and 3D) stability maps (in sections  4 and  5)  for a wide range of eccentricities [0,1] as well as a wide range of inclinations [0,$180^\circ$]. These maps enable us to access the similarities and differences between  prograde and retrograde dynamical stability.  Our conclusions about the dynamics of capture and stability in the coorbital resonance are given in section  6.

\section{Critical angles of the coorbital resonance at arbitrary inclination}
Classical expansions of the disturbing function  of the circular restricted 3-body problem (CR3BP)\footnote{We use standard Keplerian elements for the test particle's orbit: $a$ (semi-major axis), $e$ (eccentricity), $I$ (inclination), $f$ (true anomaly), 
$\omega$ (argument of pericentre), $\Omega$ (longitude of pericentre), $M$ (mean anomaly), $\lambda$ (mean longitude). The  planet's  circular orbit has semi-major axis $a_{\rm J}=1$ and mean longitude $\lambda_{\rm J}$.} have been developed to model prograde nearly coplanar  motion (relative inclination $I\approx 0$). For instance, Laplace's literal expansion of the prograde disturbing function \citep{ssdbook} involves powers of  the small parameter $\sin^2(I/2)$  hence it is invalid for  retrograde nearly coplanar motion ($I\approx180^\circ$).  
As explained in Paper I  the expansion of the retrograde disturbing function should be done in powers of the small parameter $\cos^2(I/2)$, or equivalently, by applying  the following canonical transformation to the expansion of the prograde disturbing function\footnote{This is equivalent to obtaining a retrograde configuration of mutual inclination $I>90^\circ$ from a prograde configuration of  mutual inclination $I<90^\circ$ by inverting the motion of the  planet with respect to the  star.}:
\begin{equation}
\parbox{2cm}{$I^\star=180^\circ-I$}\quad\quad 
\parbox{2cm}{$\lambda_{\rm J}^{\star}=-\lambda_{\rm J}$}\quad 
\parbox{2cm}{$\omega^\star=\omega-\pi$}\quad 
\parbox{2cm}{$\Omega^\star =-\Omega-\pi$} \  .
\label{canonical}
\end{equation}
From Eq.~\ref{canonical}  we obtain the transformations between longitudes of pericentre $\varpi^\star=\varpi-2\,\Omega$ and mean longitudes $\lambda^\star=\lambda-2\,\Omega$. When $I=180^\circ$ (planar retrograde case) the line of nodes is undefined hence $\varpi^\star=\varpi$ and $\lambda^\star=\lambda$.
 
As shown in Paper I this method allows to identify the terms associated with a $p:q$ retrograde resonance which appear at order $p+q$ in the small parameters  $e$ and $\cos(I/2)$,  in contrast with the  terms associated with a $p:q$ prograde resonance  that appear at order $p-q$ in the small parameters $e$ and $\sin(I/2)$.
In particular, when $p=q=1$ the lowest  order terms in the retrograde expansion are proportional to $e^2\cos(\lambda^\star+\lambda_{\rm J}^{\star}-2\,\varpi^\star)$ and $\cos^2(I/2)\cos(\lambda^\star+\lambda_{\rm J}^{\star}-2\,\Omega^{\star})$, while the lowest  order terms in the prograde expansion are  proportional to $\cos(\lambda-\lambda_{\rm J})$ and  $e^2\,\sin^2(I/2)\cos(\lambda-\lambda_{\rm J}-2\,\varpi+2\,\Omega)$. Moreover, written in terms of standard orbital elements, the angles with $p=q=1$ that appear at 2nd order in the retrograde expansion are $\phi^{\star}=\lambda-\lambda_{\rm J}-2\,\varpi+2\,\Omega$ and $\phi=\lambda-\lambda_{\rm J}$, associated with terms ${\cal O}(e^2)$ and ${\cal O}(\cos^2(I/2))$, respectively.  When $I=180^\circ$ there is a single angle  $\phi^{\star}$ and when $I=0$ there is a single angle $\phi$.

Expansions of the disturbing function do not converge when the semi-major ratio $\alpha\to1$ and consequently are not appropriate to model the coorbital resonance which is best studied semi-analytically \citep{Namouni1999,Namounietal1999,Nesvorny_etal2002CMDA}.
The angles $\phi$ and $\phi^\star$ appear naturally in the expansion-free expression of the averaged Hamiltonian of the 3D coorbital problem (Paper I). When $I=0$ or $I=180^\circ$  there is a single angle ($\phi$ or  $\phi^\star$), in agreement with the   analysis presented above, making the 2D (retrograde or prograde) coorbital problems integrable.  In Paper I we constructed surfaces of section of the planar CR3BP exploring the phase-space in the vicinity of the major retrograde resonances. These  showed that in 2D there are three possible retrograde coorbital modes:  
$\phi^\star$ librating around 0 which occurs at a wide range of eccentricities (mode I or R1);  $\phi^\star$ librating around $180^\circ$, at large eccentricity (mode II or R2) or small eccentricity (mode III or R3).  These resonant orbits are, as expected, associated with the equilibria of the 2D retrograde coorbital Hamiltonian (Paper I).

\begin{figure*}
\centering
 \includegraphics*[width=6.6cm]{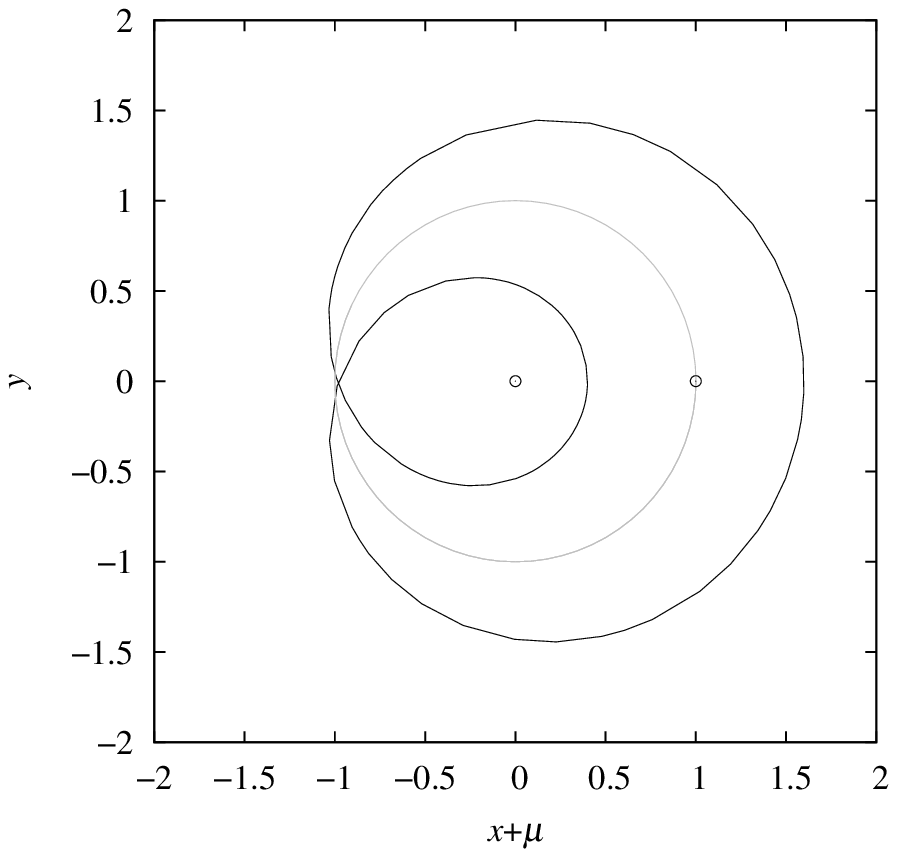}\includegraphics*[width=6.6cm]{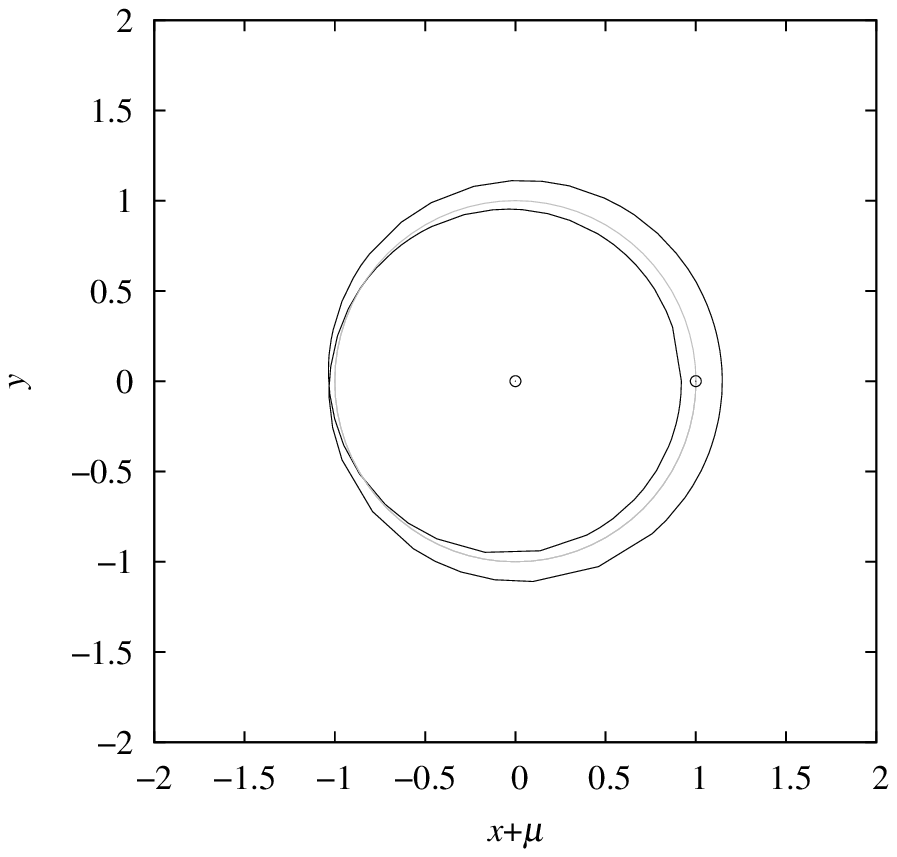}  \\     \includegraphics*[width=6.6cm]{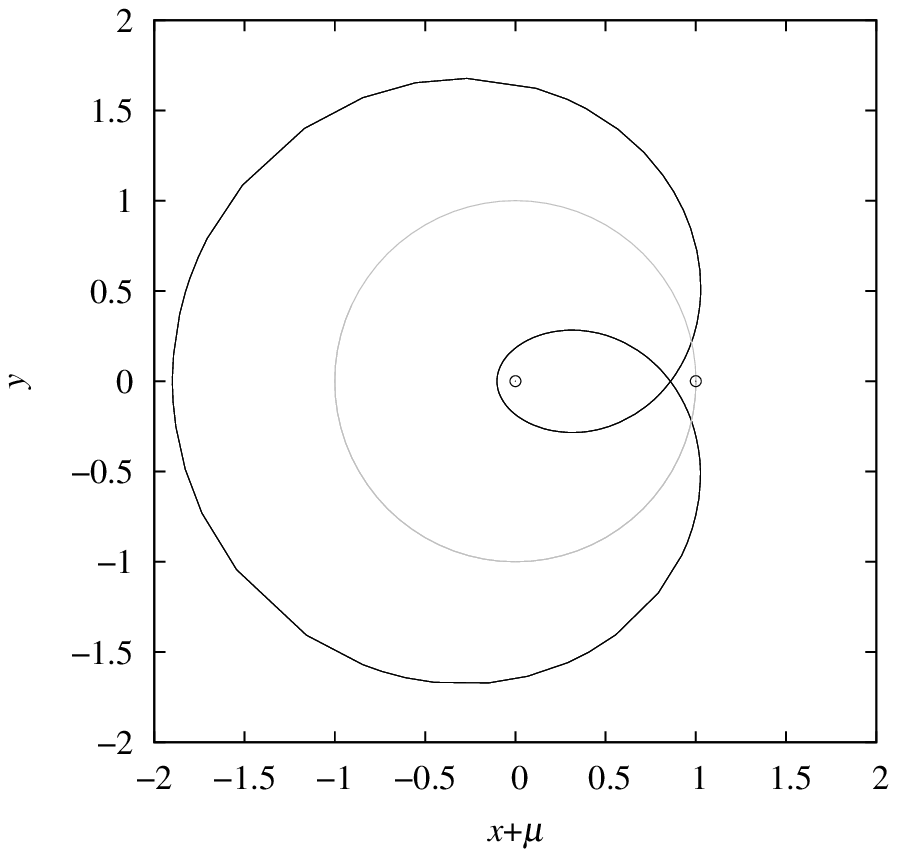}\includegraphics*[width=6.6cm]{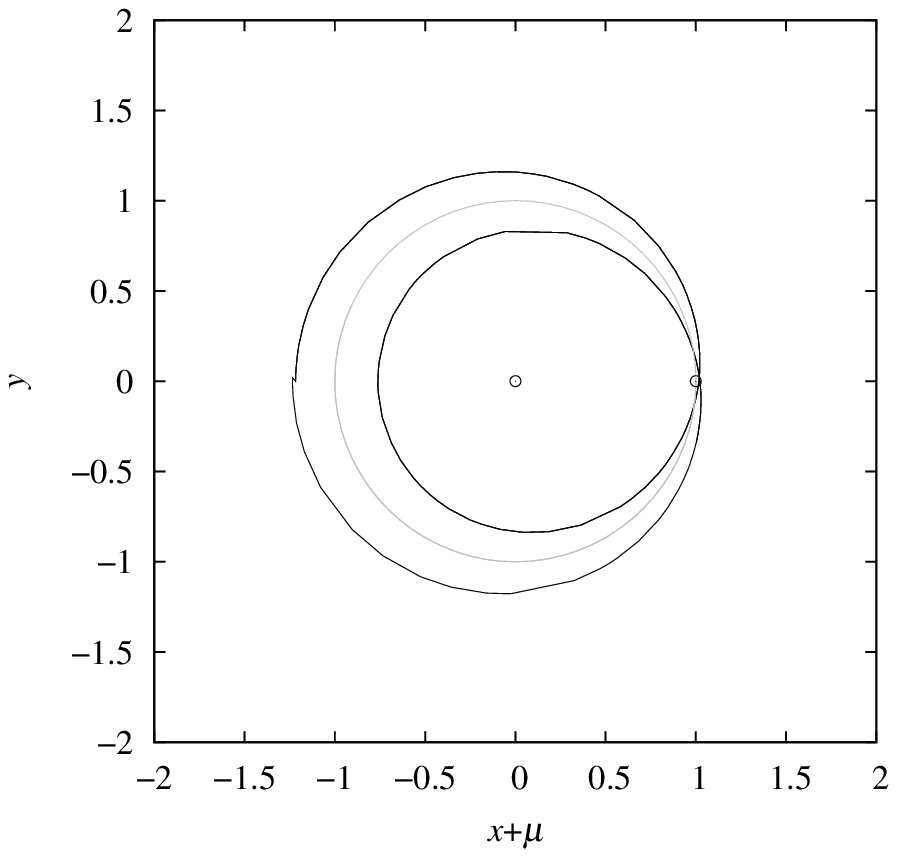}  \\   
\caption{Orbits in the planar retrograde coorbital  resonance at mass ratio $\mu=0.001$ seen in synodic frame. Initial  $\phi^\star=0$ and  $a=1.0$, $e=0.9$ (top left: mode I or R1); $a=1.0450685$, $e=0.12$   (top right: mode I or R1). Initial  $\phi^\star=180^\circ$ and  $a=1.0$, $e=0.9$ (low left: mode II or R2), $a=1.0138672$, $e=0.2$ (low right: mode III or R3). The  star (Sun) is located at $(0,0)$ and the  planet  (Jupiter) at $(1,0)$. The gray circle of unit radius helps locate the inner and outer parts of the resonant obits. The pericenter and apocenter of the orbits are the points with $y=0$ and $\dot{x}=0$. The minimum distance to the  planet of mode R3 orbit (low right panel) is larger than $0.01$ and the orbit is stable over  long integration times.
}    
\label{xy11}
\end{figure*}

Fig.~\ref{xy11} shows planar retrograde coorbitals in the  vicinity of the possible resonant centers viewed in the synodic frame. These were obtained by numerical integration of the CR3BP's equations of motion with mass ratio $\mu=0.001$ and unit distance between star and  planet. Mode R1 orbits are stable at a wide range of eccentricities, from high $e$ values  (Fig.~\ref{xy11}: top left)  limited by the collision with the  star at $e\approx 1$, down to low $e$ values (Fig.~\ref{xy11}: top right)  limited by the collision separatrix\footnote{The chaotic region due to the cumulative effect of disruptive close encounters with the  planet.} located at  pericentre and apocentre. Mode R2 orbits are only possible at high eccentricity  (Fig.~\ref{xy11}: low left) and the lower limit on the eccentricity is set by the collision separatrix located halfway between pericentre and apocentre. Mode R3 orbits only occur at small eccentricity  and are stable over long integration times despite having close encounters to the  planet halfway between pericentre and apocentre.
There are three possible 2D prograde coorbital modes: $\phi$  librating around 0 (retrograde-satellite orbits\footnote{ Also known as quasi-satellite orbits \citep{Mikkola_etal2004}. These have clockwise motion around the planet in the synodic frame and are not true satellites since the orbits are bound to the star.}: RS), $\pm60^\circ$ (tadpole orbits: T) or $180^\circ$ (horseshoe orbits: H).

Coorbital motion in 3D has been studied only in the case of prograde inclination \citep{Namouni1999,Namounietal1999,Nesvorny_etal2002CMDA}. Retrograde planar  coorbital modes have been recently discovered (Paper I) hence it is expected  that at least some of these also exist when $90^\circ<I<180^\circ$. The purpose of this article is to study numerically the case of retrograde inclination and to relate the retrograde coorbital modes that appear in 2D and 3D with recent results regarding capture of retrograde orbits in the coorbital resonance (Paper II). 

\section{Coorbital capture of retrograde orbits: 2D vs 3D}
Capture of a drifting particle in the Jupiter-Sun three-body problem was examined in Paper I in terms of statistics as function of the particle's initial eccentricity, inclination and the planet's eccentricity. 
In the following we examine closely capture in the retrograde coorbital resonance and show that its dynamics is richer than previously thought. In particular, extrapolating the workings of the capture process from the planar problem is not adequate to understand the dynamics in three dimensions. 

\begin{figure*}
\centering
\includegraphics[width=.9\textwidth]{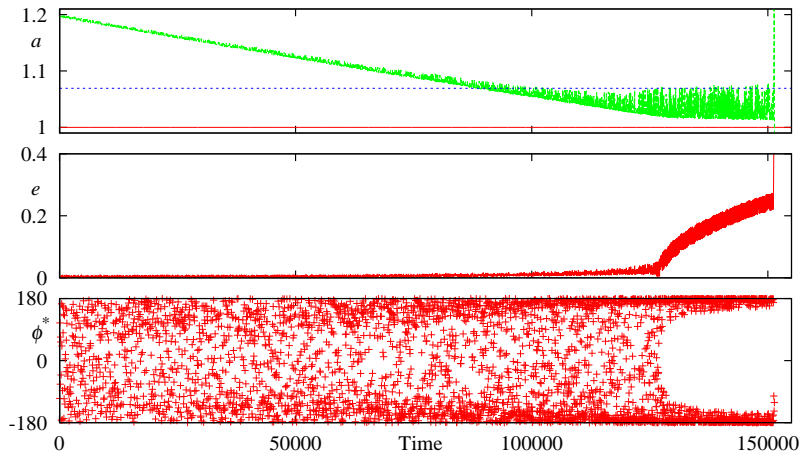} \\[-20mm]
\caption{Capture in the planar problem. The particle's semi-major axis $a$, eccentricity $e$ and resonant argument $\phi^\star= \lambda + \lambda_{\rm J}-2\varpi$ are shown as a function of time in units of the planetary period. In the upper panel, the red solid curve is the planet's semi-major axis and the dotted blue line is the planet's  Hill radius. }
\label{fig2}    
\end{figure*}

We first start by recalling how capture proceeds in the planar retrograde problem and consider a particle that drifts from a semi-major axis  equal to 1.2 units of Jupiter's orbit semi-major axis with both bodies on  initially circular orbits.  Orbital drift is modeled by a drag force of the type $-k {\bf v}$ where $k$ is the friction coefficient and $ {\bf v}$ the particle's velocity. This implies a semi-major axis function $a(t) =a(0) \exp (-2kt)$ with a characteristic drift time $(2k)^{-1}$ that we set to  $8\times 10^{5}$ planetary orbital periods. Resonance crossing is therefore adiabatic  and the initial semi-major axis close to the planet is set to maximize the likelihood of capture in the coorbital resonance. For the planar configuration, it is simpler to use the convention adopted by \cite{Morais&Giuppone2012MNRAS} hence we integrate the 2D equations of motion with negative mean motion of the planet.  The planar coorbital retrograde resonance argument is $\phi^\star= \lambda + \lambda_{\rm J}-2\varpi$ where $\dot{\lambda}_{\rm J}<0$ due to the planet's negative mean motion, $\lambda$ and $\varpi$ are the standard mean longitude and  longitude of perihelion.  Figure~\ref{fig2} shows that capture in the coorbital mode with  center $\phi^\star=180^\circ$ (mode R3) occurs at around 1.3$\times 10^5$ orbital periods when the particle is {\it inside} Jupiter's Hill sphere of radius $0.0693$.   According to Paper II, 2D initial circular orbits are captured with unit probability in mode R3. Close approaches to the planet's collision separatrix \citep{Namouni1999} result in large semi-major axis and eccentricity perturbations that ultimately eject the particle from the planet's vicinity. 

\begin{figure*}
\centering
\includegraphics[width=.9\textwidth]{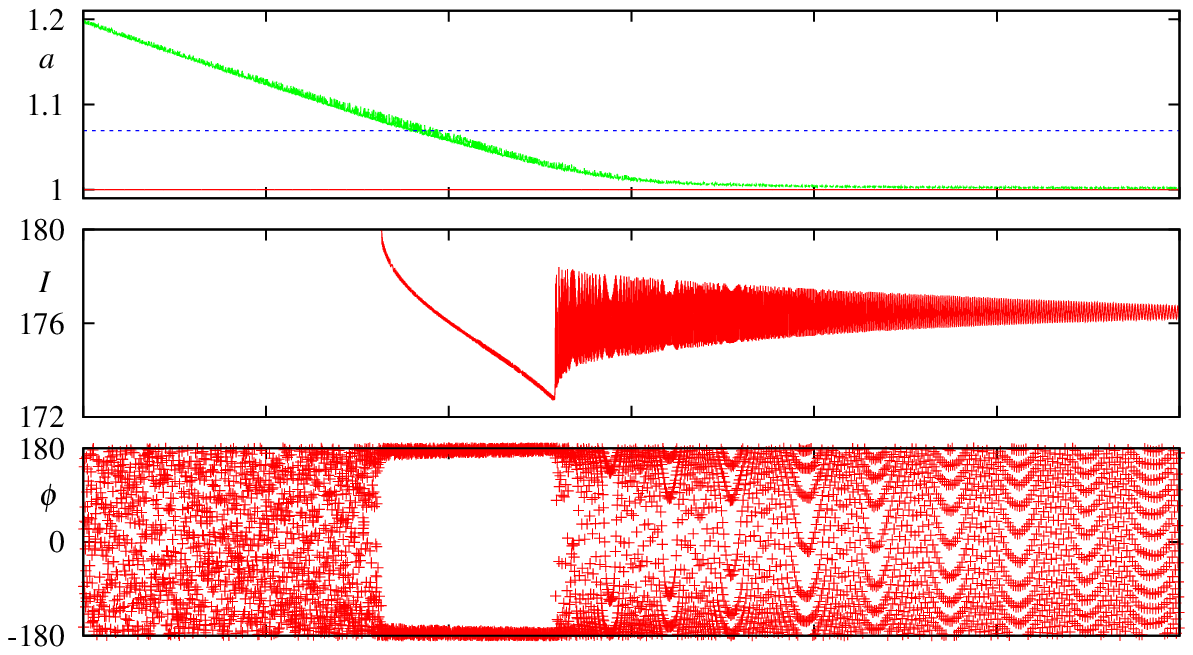} \\[-20mm]
\includegraphics[width=.9\textwidth]{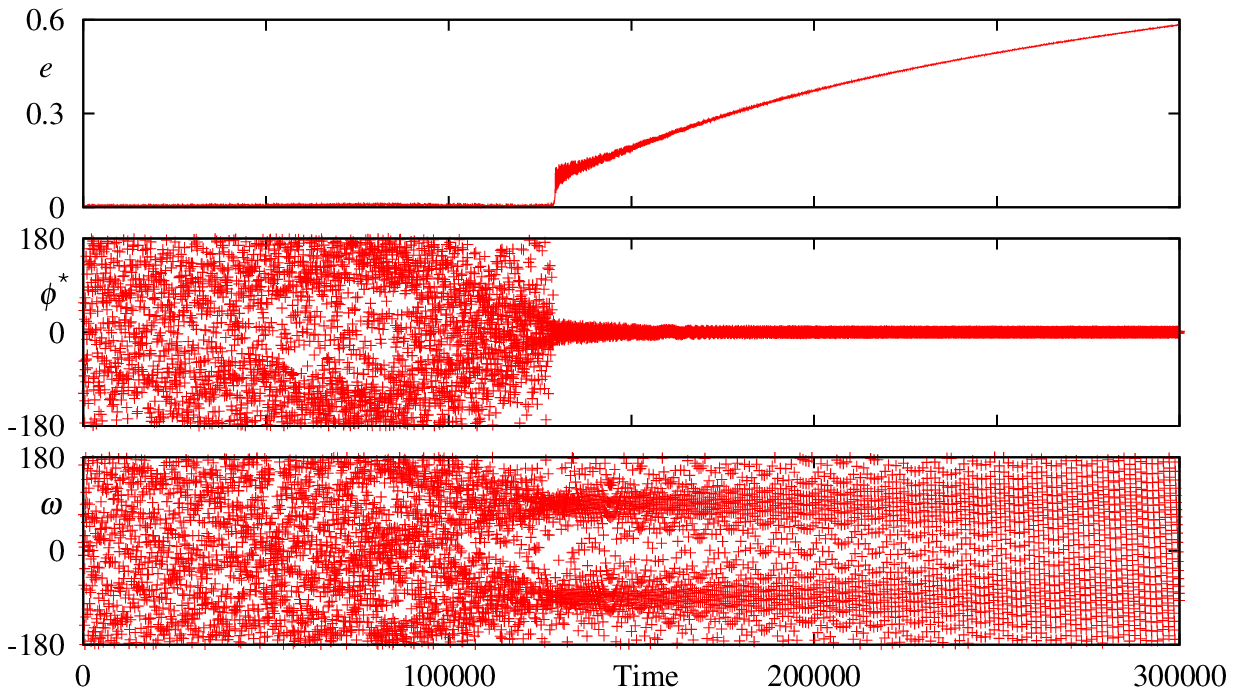} \\[-20mm]
\caption{Capture in the three-dimensional problem with initial inclination $179.999^\circ$. The particle's semi-major axis $a$, eccentricity $e$ and inclination  are shown as a function of time in units of the planetary period. Also shown are the possible resonant angles: $\phi=\lambda^\star-\lambda_{\rm J}+2\Omega$, $\phi^\star=\lambda^\star-\lambda_{\rm J}-2\varpi^\star$. In the upper panel, the red solid curve is the planet's semi-major axis and the dotted blue line is the planet's Hill radius. }
\label{fig3}
\end{figure*}
\begin{figure*}
\centering
\includegraphics[width=.8\textwidth]{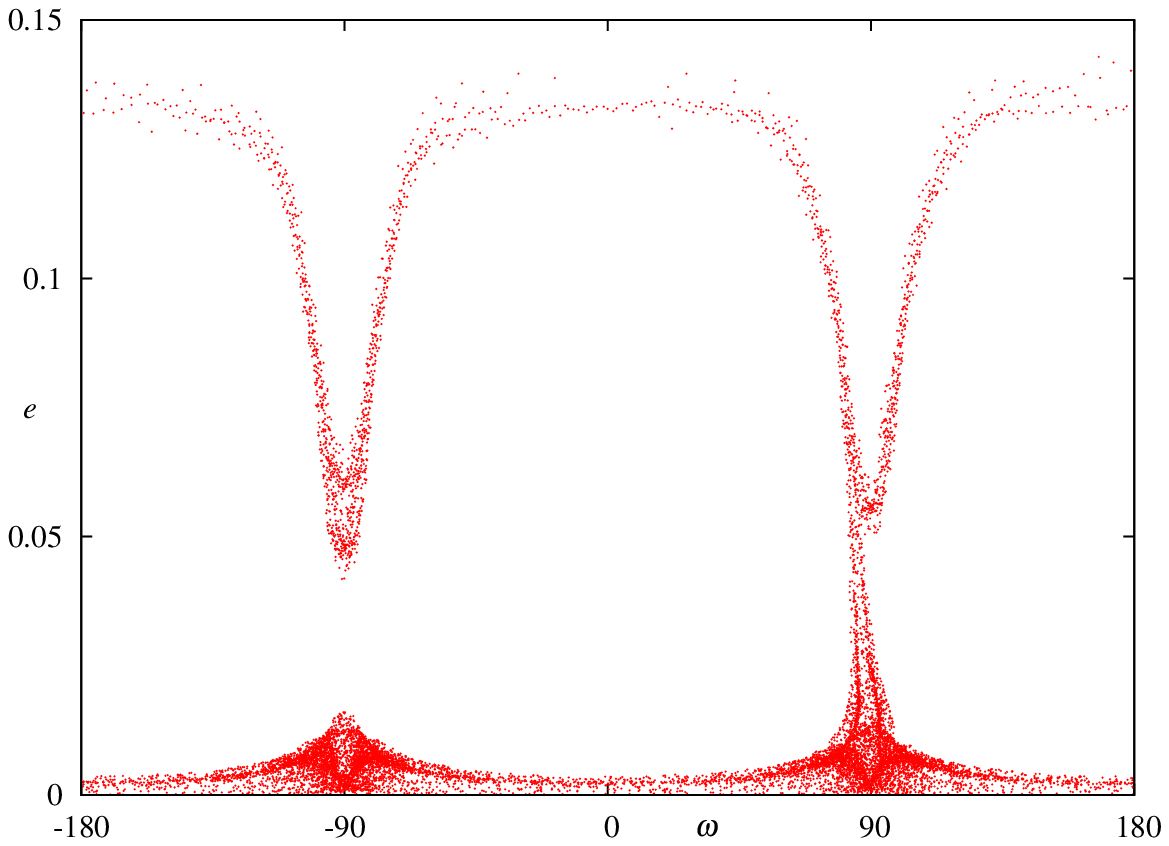} \\
\caption{Kozai-Lidov resonance stage (between $1.25 \times 10^5$ and $1.3\times 10^5$) from Fig.~\ref{fig3} (see text for details).}
\label{fig4}
\end{figure*}

In order to reveal the different dynamical processes at work with respect to the planar problem
we repeat this simulation integrating the 3D equations of motion where the mean motion of the planet is positive and the the test particle has initial inclination  $179.999^\circ$. In three dimensions, a retrograde orbit has inclination $180^\circ>I>90^\circ$ and the convention of Paper I  and Sect.~2 for the angles is used.  The results are shown in Figure~\ref{fig3}. Unlike the planar configuration, capture occurs {\it outside} the Hill sphere (of radius $R_{\rm H}=0.0693$) at around 8.1$\times 10^4$ orbital periods in the coorbital mode with argument $\phi=\lambda^\star-\lambda_{\rm J}+2\Omega$  and center $\phi=180^\circ$ (mode R4) . As explained in Sect.~2, when written in terms of the standard orbital elements, the argument $\phi=\lambda-\lambda_{\rm J}$ is that of the standard prograde coorbital resonance.  However, for retrograde motion the resonance behaves exactly like an inclination resonance witness the steady decrease (increase) of (relative) inclination whereas the eccentricity remains constant. Capture in  coorbital mode R4 lasts about  5 $\times 10^4$ orbital periods  until the Kozai-Lidov resonance is triggered.  In Figure~\ref{fig4}, we show how the Kozai-Lidov resonance affects 3D capture in the time interval between $1.25 \times 10^5$ and $1.3\times 10^5$. Since the initial eccentricity is zero, $\omega$ circulates near the Kozai-Lidov centers at  $\pm 90^\circ$ and $e\approx 0.01$. As the semi-major axis slowly decreases  the topology of the secular phase-space changes and the particle  gets trapped at the $\omega=-90^\circ$ and later at the $\omega=90^\circ$ centers (evidenced by the higher density of points in those regions).  The libration island at $\omega=90^\circ$ gets displaced to higher eccentricity and eventually the orbit reaches  the vicinity of the Kozai-Lidov separatrix  associated with the $\omega=0$ and $\omega=180^\circ$ islands. The angle $\omega$ circulates again but now the eccentricity oscillates between 0.05 and 0.13 allowing capture  in the coorbital mode with  argument $\phi^\star=\lambda^\star-\lambda_{\rm J}-2\varpi^\star$ and center $\phi^\star=0$ (mode R1).
The particle remains trapped in  coorbital mode R1 while the eccentricity increases steadily, the typical behavior of an eccentricity resonance. Collisions are unlikely until the particle's orbit reaches unit eccentricity.\footnote{If all bodies are treated as point-like objects and evolution is allowed to continue, no collision occurs. Instead orbital reversal takes place \citep{YuandTremaine}.}  

The presence of the Kozai-Lidov mechanism during mean motion resonance capture is a process that was observed in resonance capture at arbitrary inclination in our intensive ($\sim 600\, 000$) simulations of Paper II.  In order to illustrate further its workings, we present another 3D simulation  with identical initial conditions except for the initial inclination set to $175^\circ$ and a radial drift 10 times slower than previously. We increased relative inclination so as to give a stronger effect to inclination resonances. Similarly we slowed down the semimajor axis drift so as to allow the particle to explore a larger portion of the secular phase space. Figure 5 shows the evolution of the orbital elements and resonant angles from the capture in mode R4 ($\phi=180^\circ$) to the capture in mode R1 ($\phi^\star=0$). New features appear such as the temporary exit from mode R4 corresponding to brief Kozai-Lidov oscillation around $\pm 90^\circ$ as well as the long capture phase in the $\omega=0$ Kozai-Lidov resonance preceding capture in mode R1. 
Capture in the inclination-type retrograde resonance  with centre $\phi=180^\circ$ (mode R4) has a relatively shorter  duration (with respect to drift time) when  initial inclination is decreased. We have checked that the three-stage capture mechanism is  generic and  independent of the semi-major axis drift rate as long as  this is much slower than the libration period to allow for an adiabatic resonance crossing. Outwardly migrating particles with initially circular orbits interior to the planet do not get captured in the coorbital zone whether the retrograde configuration is plane or slightly inclined.  

\begin{figure*}
\centering
\includegraphics[width=.9\textwidth]{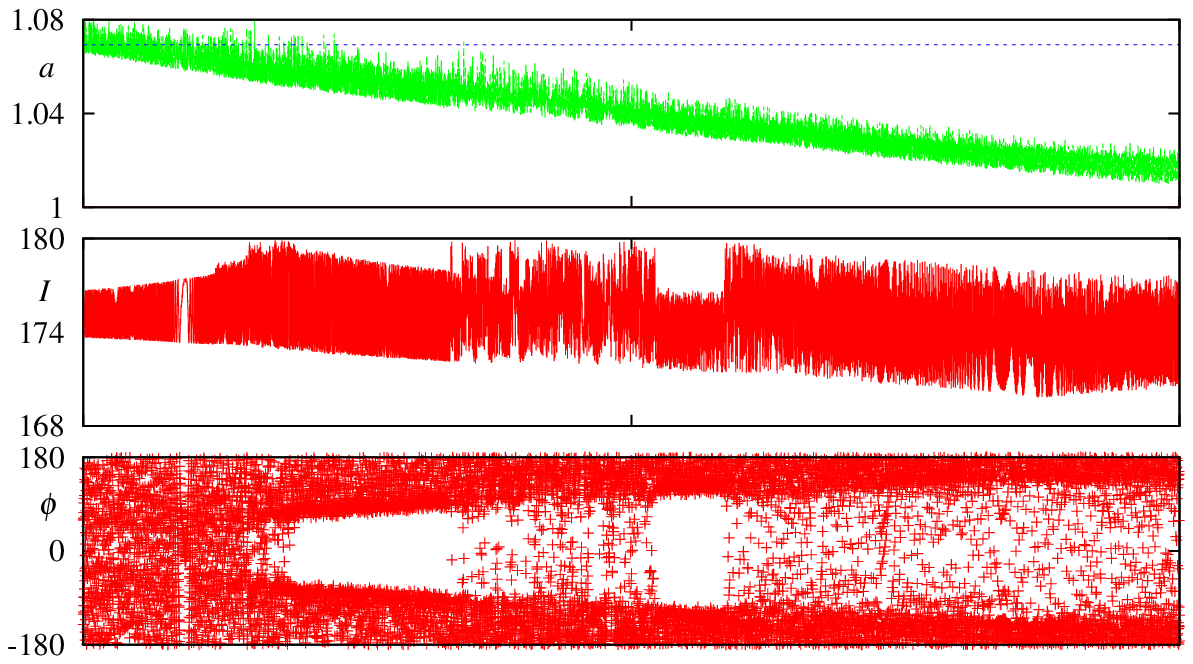} \\[-20mm]
\includegraphics[width=.9\textwidth]{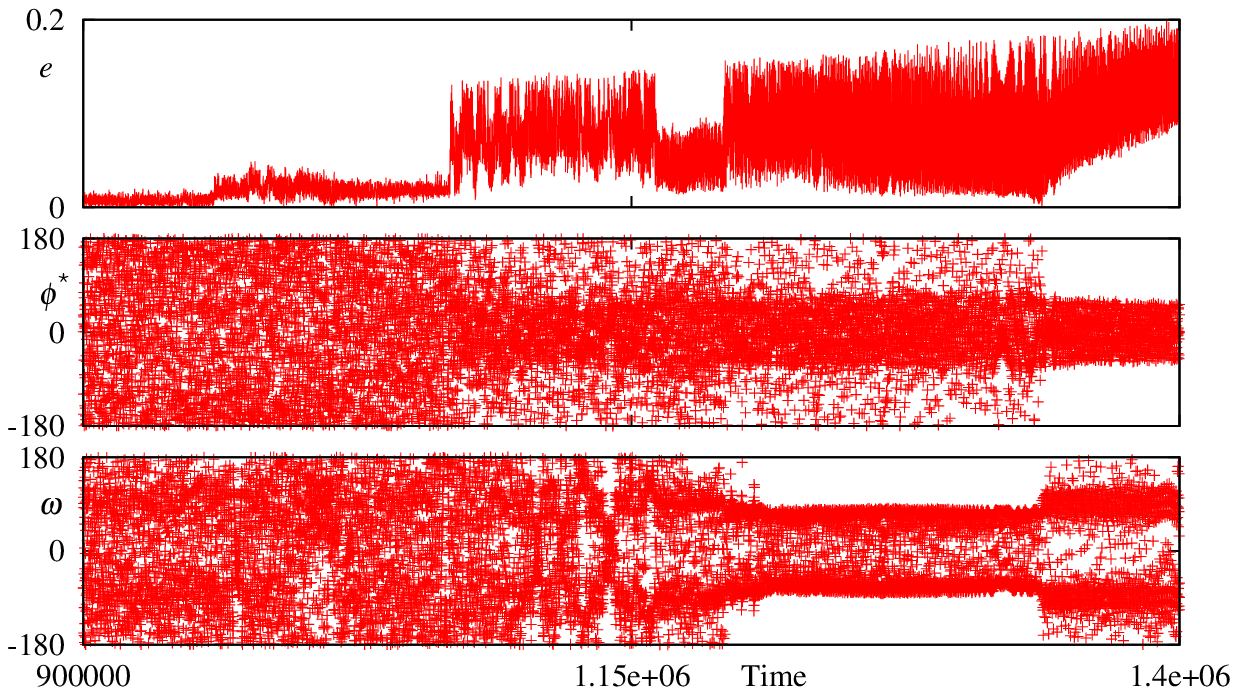} \\[-20mm]
\caption{Capture in the three-dimensional problem as in Figure 3 but with initial inclination $175^\circ$ and radial drift 10 times slower than in previous simulations. }
\label{fig5}
\end{figure*}

\section{Stability analysis for 2D configurations}
Stability analysis is an important tool to understand how capture occurs in the three-body problem. By charting parameter space, one is able to identify the possible  dynamical routes that the particle's orbit may take to enter resonance.  To this end, we numerically integrate the equations of motion of the circular restricted three-body problem consisting of a Sun-mass star, a Jupiter-like planet with a mass ratio $\mu=0.001$ and a massless particle,  together with the variational equations and MEGNO equations  \citep{Cincotta2000,Gozdziewski2003A&A} for $5\,\times 10^4$ periods using a Burlisch-Stoer method with double precision arithmetics and tolerance $10^{-14}$.  The mean MEGNO $\langle Y\rangle$ converges to $2$ for regular orbits and increases at a rate inversely proportional to  Lyapunov's time for chaotic orbits  \citep{Cincotta2000,Gozdziewski2003A&A}. We set the maximum mean MEGNO $\langle Y\rangle$ value for chaotic orbits at $8$  in order to present stability maps with a high contrast between the regular and chaotic regions.
The integration of each test particle's orbit is stopped before  $5\,\times 10^4$ periods if collision or escape occur: a collision occurs  when the minimum distances to the primary or planet are less than $0.005$ or $0.0005$, respectively (the unit distance is the Sun-Jupiter semi-major axis as in the previous section). These values correspond approximately to the solar and jovian radii. Escape occurs when the barycentric distance exceeds the value 3.

\subsection{Retrograde orbits}
\begin{figure*}
\centering
 \includegraphics[width=0.9\textwidth]{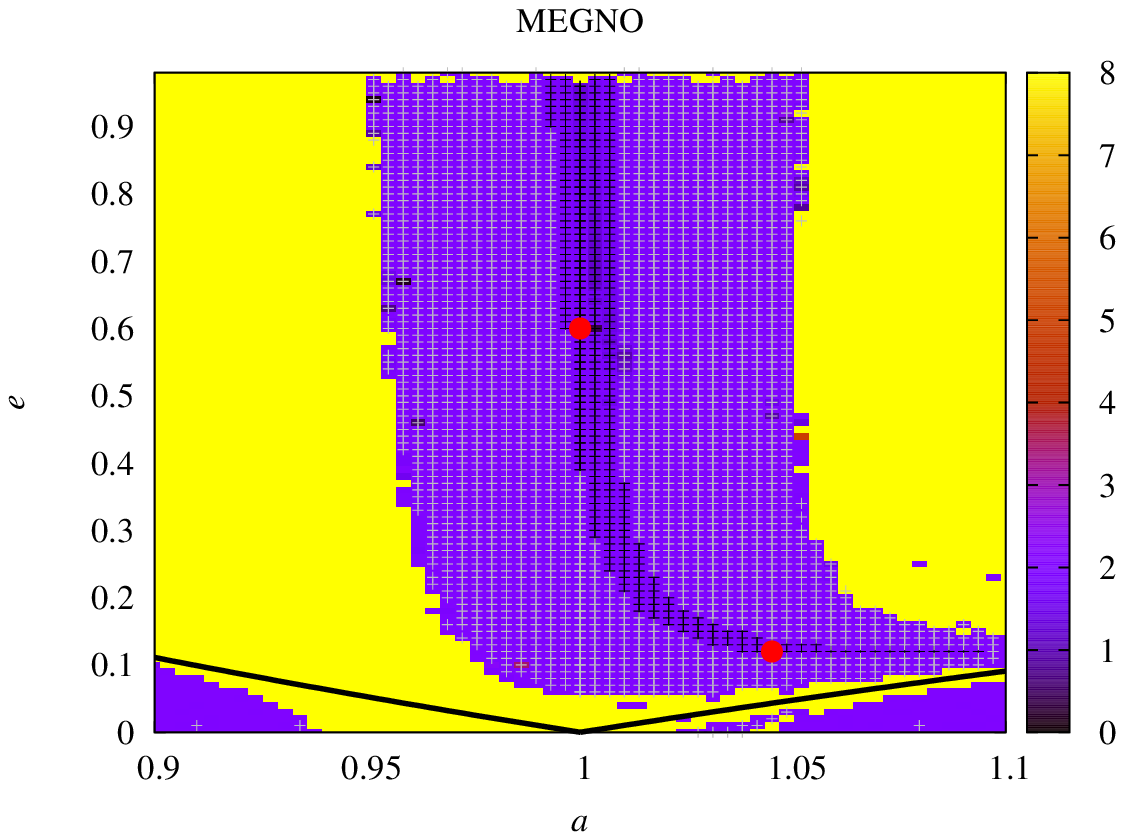} 
 \includegraphics[width=0.9\textwidth]{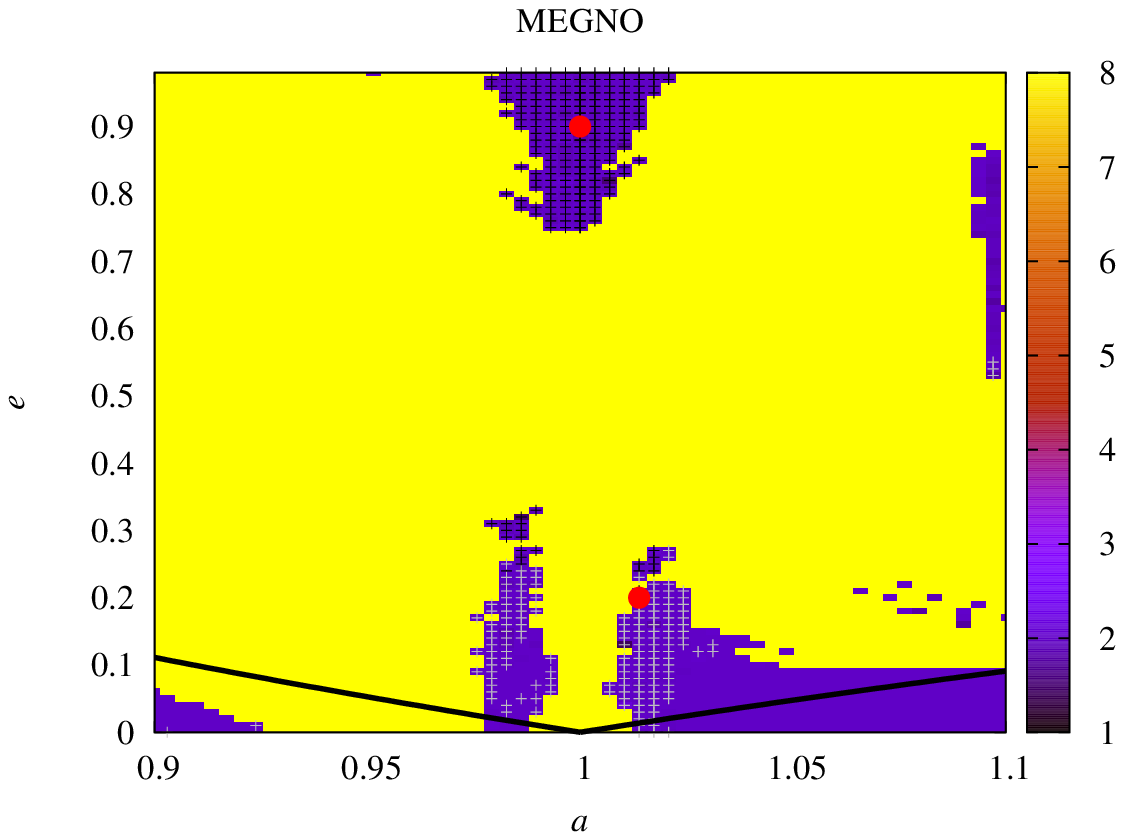}
\caption{Stability map for the planar retrograde coorbital resonance  with  $\phi^{\star}=0$ (top panel) or $\phi^{\star}=180^{\circ}$ (low panel) when $\mu=0.001$. The symbol $+$ marks initial conditions such that $\phi^{\star}$ librates with amplitude less than $50^{\circ}$ (black $+$) or larger than $50^{\circ}$ (gray $+$). The black lines represent collision at pericentre (left hand side) or apocentre (right hand side). The red circles indicate initial conditions of orbits shown in Fig~\ref{xy11}.}
\label{megnoR11}       
\end{figure*}

The test particle's initial conditions for the coorbital retrograde resonance are: inclination $I=180^\circ$, semi-major axis in $(1-2\,R_{\rm H},1+2\,R_{\rm H})$ varying at steps $0.05\,R_{\rm H}$, eccentricity in $(0,1)$ varying at steps $0.01$, angles $\Omega=\omega=0$ where $\omega=\varpi-\Omega$, and $\phi^{\star}=\lambda-\lambda_{\rm J}-2\,\omega=0,180^\circ$.  In Fig.~\ref{megnoR11} we show the MEGNO maps for configurations with $\phi^{\star}=0$ (top panel) and $\phi^{\star}=180^\circ$ (low panel).

Coorbital  mode R1 where $\phi^\star$ librates around 0 occurs for eccentricity $e\gtrsim 0.1$.
Close encounters with the planet occur at pericentre and apocentre for small  eccentricity orbits  (Fig.~\ref{xy11}: top right) hence the pericentric and apocentric collision lines limit the resonance borders in Fig.~\ref{megnoR11} (top panel). 
The apocentric collision line (right hand side) is surrounded by a thin chaotic layer which extends to $a>1.1$, separating the outer circulating region from the librating region (mode R1). 

The coorbital modes where $\phi^\star$ librates around $180^\circ$ occur at   eccentricity $e\gtrsim0.7$ and $1.02 \gtrsim a\gtrsim 0.98$ (mode R2), or eccentricity $e\lesssim0.3$ in the semi-major axis ranges  $0.99\gtrsim a \gtrsim 0.98$ and $1.03\gtrsim a \gtrsim1.01$ (mode R3). The orbits in the vicinity of the resonance centers in Fig.~\ref{megnoR11} (low panel)  are close to the collision separatrix located midway between pericentre and apocentre (Fig.~\ref{xy11}: low right) but collisions with the planet are avoided.  Repeated close encounters with the planet significantly disturb the osculating orbital elements which implies that these are not adequate to identify the resonance center at $\phi^\star=180^\circ$, the method of surfaces of section being more appropriate for this task (Paper I).

The resonance regions in Fig.~\ref{megnoR11}  are surrounded by  chaotic layers except for low eccentricity outer orbits with $\phi^\star$ librating around $180^\circ$ (mode R3 in the low panel).  This feature explains why circular planar retrograde orbits migrating from the outer region towards the planet are always captured in coorbital mode R3  while those migrating  from the inner region are never captured as seen in section 3.

\subsection{Prograde orbits}
\begin{figure*}
\centering
  \includegraphics[width=0.9\textwidth]{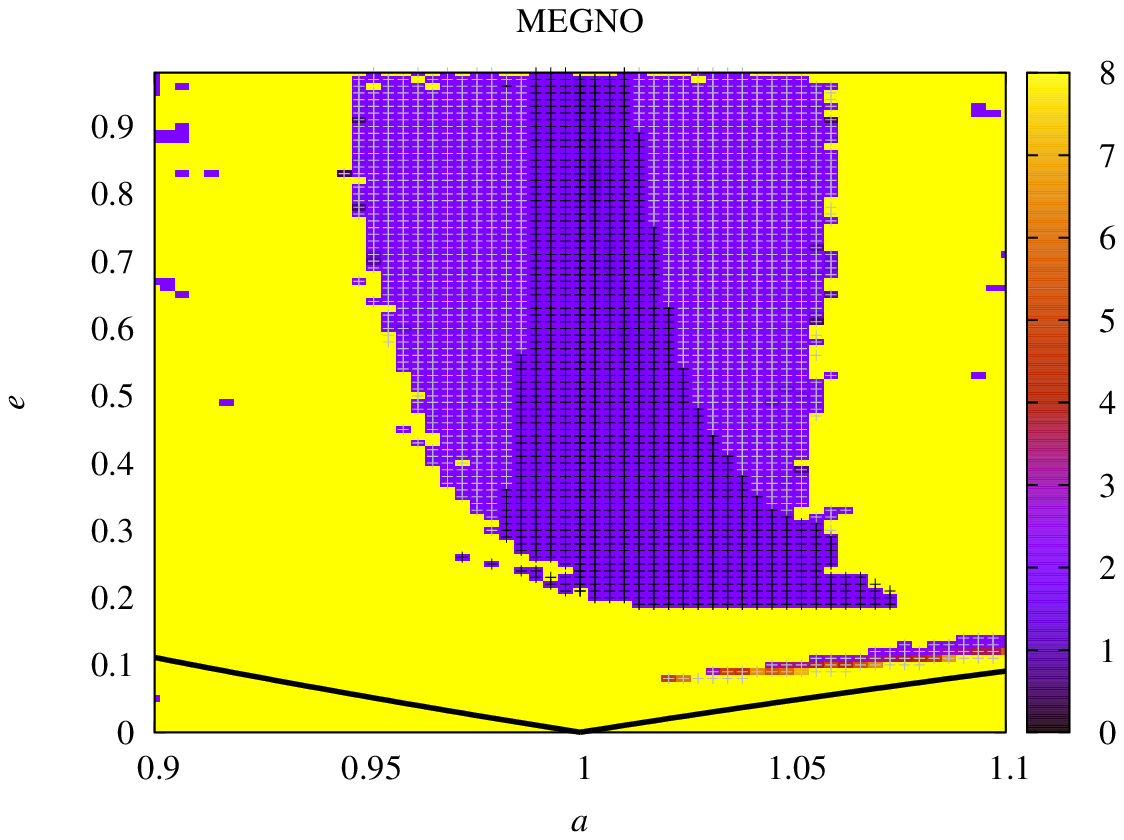} 
  \includegraphics[width=0.9\textwidth]{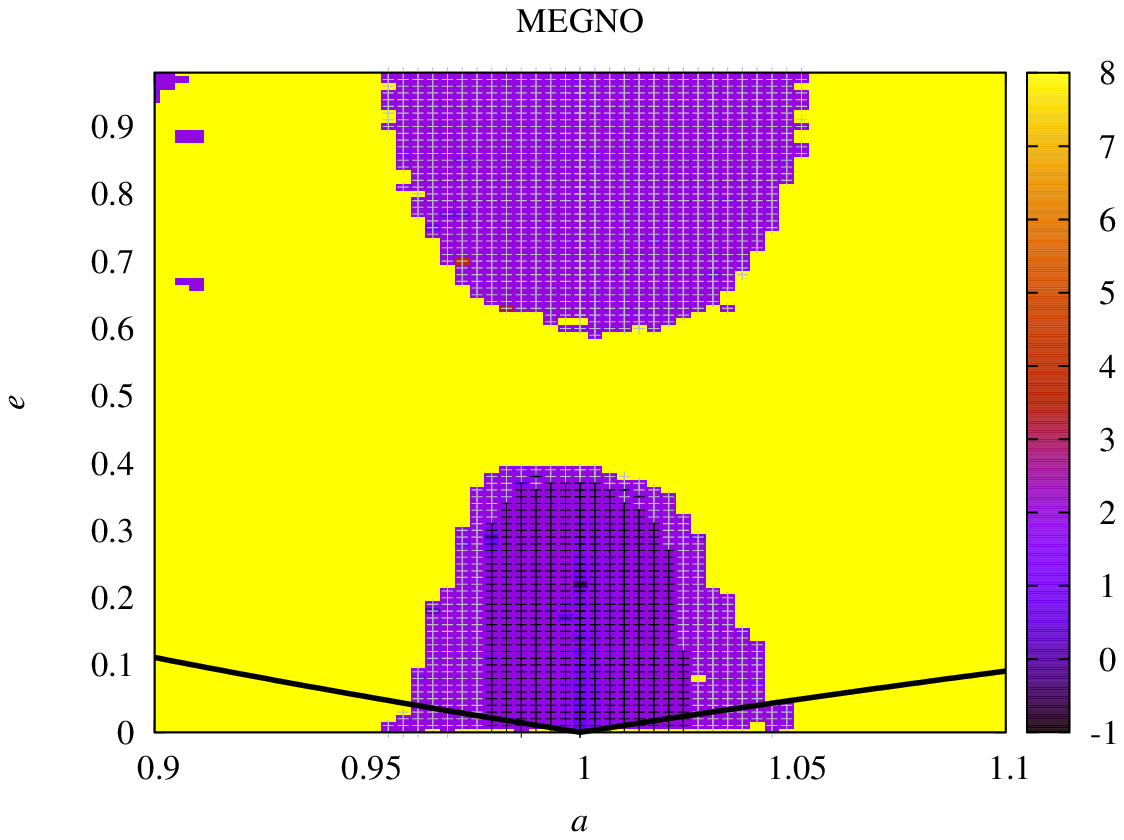}
  \includegraphics[width=0.9\textwidth]{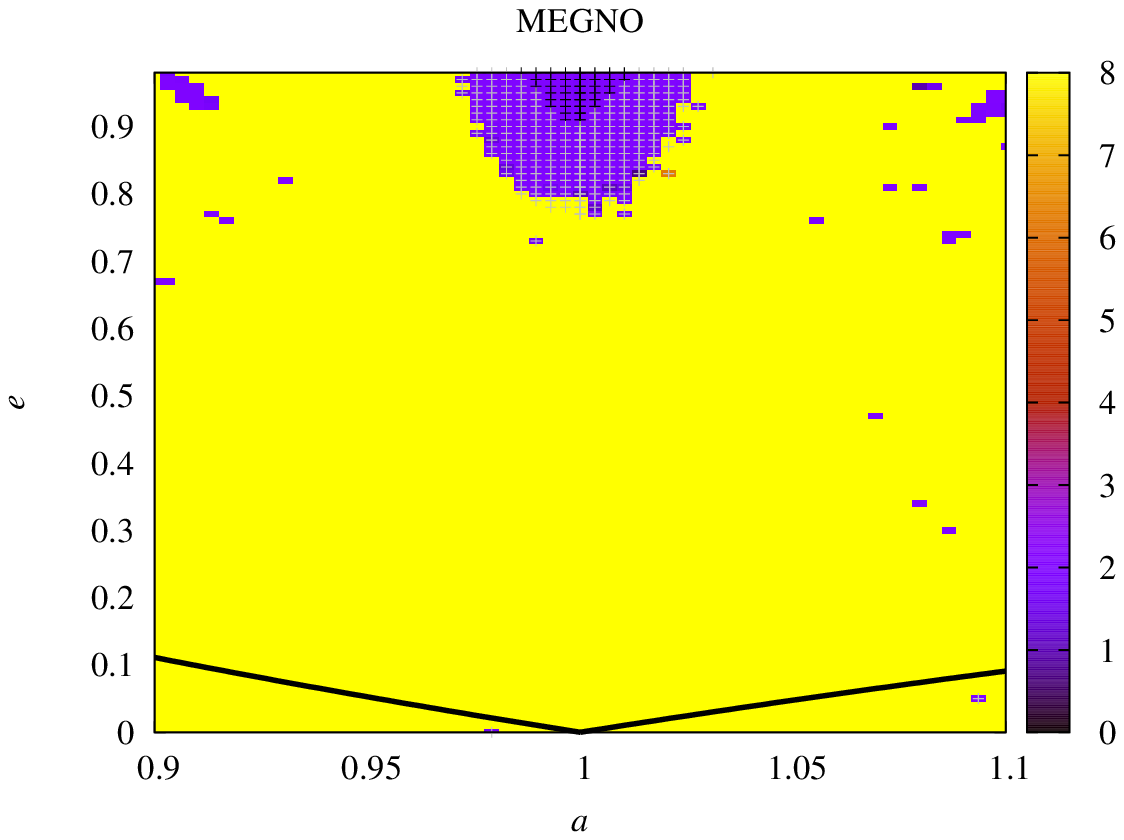}
\caption{Stability map for the planar prograde coorbital resonance  with  $\phi=0$ (top panel), $\phi=60^{\circ}$ (mid panel) or $\phi=180^{\circ}$ (low panel) when $\mu=0.001$. The symbol $+$ marks initial conditions such that $\phi$ librates with amplitude less than $50^{\circ}$ (black $+$) or larger than $50^{\circ}$. The black lines represent collision at pericentre (left hand side) or apocentre (right hand side).}
\label{megnoP11}       
\end{figure*}

 The test particle's initial conditions for the coorbital prograde resonance are identical to the retrograde case except for the inclination $I=0$, and  $\phi=\lambda-\lambda_{\rm J}=0,60^\circ,180^\circ$.  In Fig.~\ref{megnoP11} we show the MEGNO maps for configurations with $\phi=0$ (top panel), $\phi=60^\circ$ (mid panel) and $\phi=180^\circ$ (low panel) . 

Fig.~\ref{megnoP11} (top panel) shows region of libration of RS (retrograde-satellite) orbits ($\phi$ librates around 0) at $e\gtrsim 0.2$. Fig.~\ref{megnoP11} (mid panel) shows regions of libration of T (tadpole) orbits ($\phi$ librates around $\pm60^\circ$) at  $e\lesssim0.4$ and RS orbits at $e\gtrsim0.6$. Fig.~\ref{megnoP11} (low panel) shows that H (horseshe) orbits ($\phi$ librates around $180^\circ$) are only possible at $e\gtrsim0.8$ when the mass ratio is $\mu=0.001$.

All resonance regions in Fig.~\ref{megnoP11}  are surrounded by thick chaotic layers which extend over the entire range of eccentricities.  Therefore, planar prograde orbits migrating towards the planet  cannot be smoothly captured in the coorbital resonance as seen in Paper II. Instead, temporary capture due to chaotic scattering may occur.

\section{Stability analysis for 3D configurations}
The three-dimensional coorbital resonance is much richer than its planar counterpart \citep{Namouni1999,Namounietal1999}. In addition to the prograde corbital modes, tadpole (T), retrograde satellite (RS) and horseshoe (H),  retrograde satellites combine with tadpoles or horseshoes to form  the compound orbits, H-RS, RS-H, T-RS, RS-T and T-RS-T. The third dimension also  influences significantly orbital nature in another way. Whereas proper eccentricity is a secular constant in the planar configuration, it is subject to (possibly large) secular time variations along with the inclination giving rise to orbital transitions between the various libration modes as well as the appearance of the Kozai-Lidov resonance. 

Owing to this particularly complex nature of the three-dimensional coorbital problem, in order to understand the basic features of resonance capture we will not attempt to identify coorbital mode transitions or  compound orbits. 
We start our exploration by setting the particle's semi-major axis to unity. Eccentricity and inclination are chosen in a grid with $e$ between 0 and $0.99$ with a step of $0.01$ and   $I$  between $5^\circ$ and $180^{\circ}$ with a step of $5^\circ$. The initial angles were\footnote{The 1:1 resonance hamiltonian is invariant with respect to the transformation $(\phi,-\omega)\to(-\phi,\omega)$. The CR3BP  is invariant with respect to the transformation $\omega\to\omega+180^{\circ}$ and with respect to changes in $\Omega$.} $\phi=0,60^{\circ},180^{\circ}$ and  $\omega=0,90^{\circ}$. In Table 1 we show the initial values of $\phi^{\star}=\phi-2\,\omega$ for each $\phi$ and $\omega$. The choice of initial angles $\phi$ and $\phi^{\star}$ is set to cover the coorbital configurations of the planar problem that we explored earlier. From Fig.~\ref{megnoP11}  prograde coorbitals  are those in the RS mode ($\phi\sim0$); T mode ($\phi\sim\pm60^\circ$); H mode ($\phi\sim180^\circ$). From Fig.~\ref{megnoR11} retrograde coorbitals at $a=1$ are: mode R1 ($\phi^{\star}\sim0$); mode R2 ($\phi^{\star}\sim180^\circ$). We set initially $\omega=0,90^\circ$ as  these are the possible Kozai-Lidov libration centers.

\begin{table}[h]
\centering
\begin{tabular}{|l||l|l|}
\hline
        & $\omega=0$ & $\omega=90^\circ$ \\ \hline \hline
$\phi=0$ & $\phi^{\star}=0$ & $\phi^{\star}=180^\circ$ \\ \hline
$\phi=60^\circ$ & $\phi^{\star}=60^\circ$ & $\phi^{\star}=240^\circ$ \\ \hline
$\phi=180^\circ$ & $\phi^{\star}=180^\circ$ & $\phi^{\star}=0$ \\ \hline        
\end{tabular}
\caption{ Initial angles  $\phi$, $\phi^\star$, $\omega$ for simulations shown in Figs.~6\&7.}
\end{table}
                  
For each initial condition in our grid, we computed a chaos indicator (mean MEGNO $\langle Y\rangle$) and the libration amplitudes of the angles $\phi$, $\phi^{*}=\phi-2\,\omega$ and  $\omega$ that serve to identify resonant orbits. 

\begin{figure*}
\centering
 \includegraphics[width=0.85\textwidth]{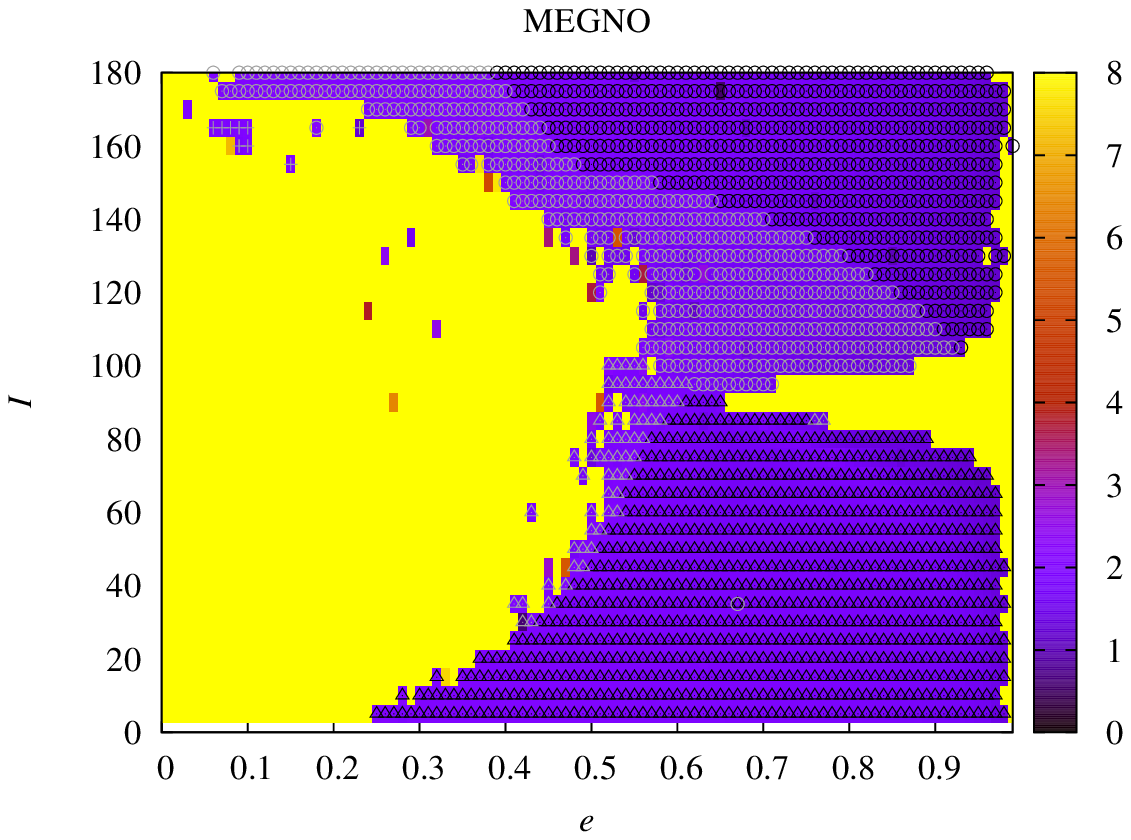}  
 \includegraphics[width=0.85\textwidth]{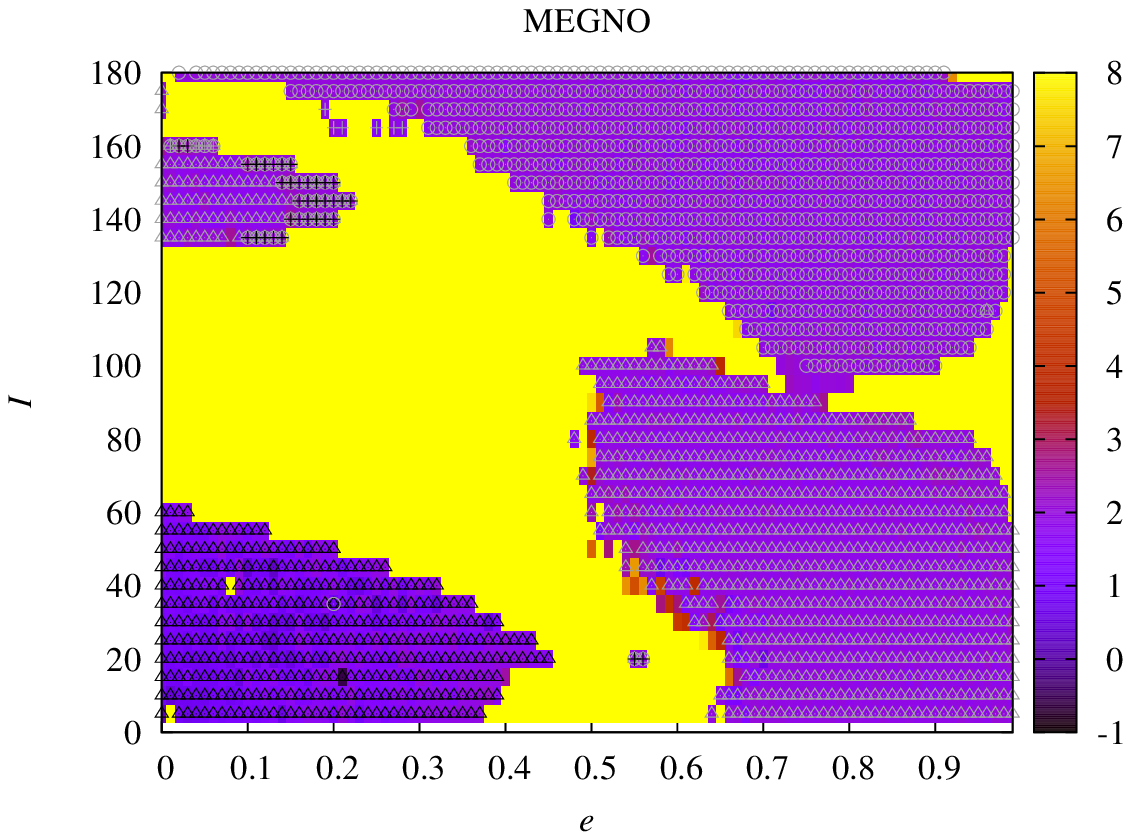}   
  \includegraphics[width=0.85\textwidth]{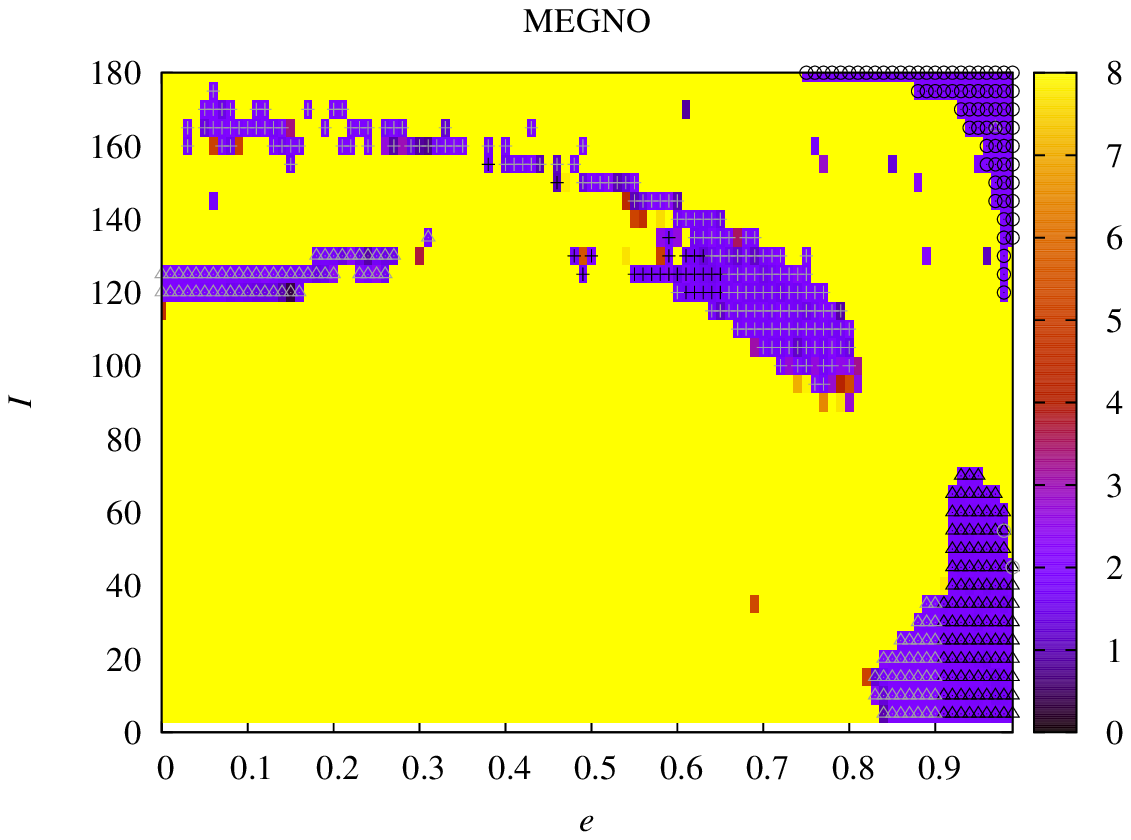}  
\caption{Stability maps for coorbital resonance in CR3BP: initial conditions $a=1$, $\omega=0$ and $\phi=0$ (top panel), $\phi=60^\circ$ (mid panel), $\phi=180^\circ$ (low panel). The symbol $+$ marks initial conditions such that $\omega$ librates with amplitude less than $50^{\circ}$ (dark $+$) or larger than $50^{\circ}$ (light $+$). The  triangles (circles) mark initial conditions such that $\phi$ ($\phi^\star$) librate with amplitude less than $50^{\circ}$ (black) or larger than $50^{\circ}$ (gray).}    
\label{megno_w0}
\end{figure*}

\begin{figure*}
\centering
  \includegraphics[width=0.85\textwidth]{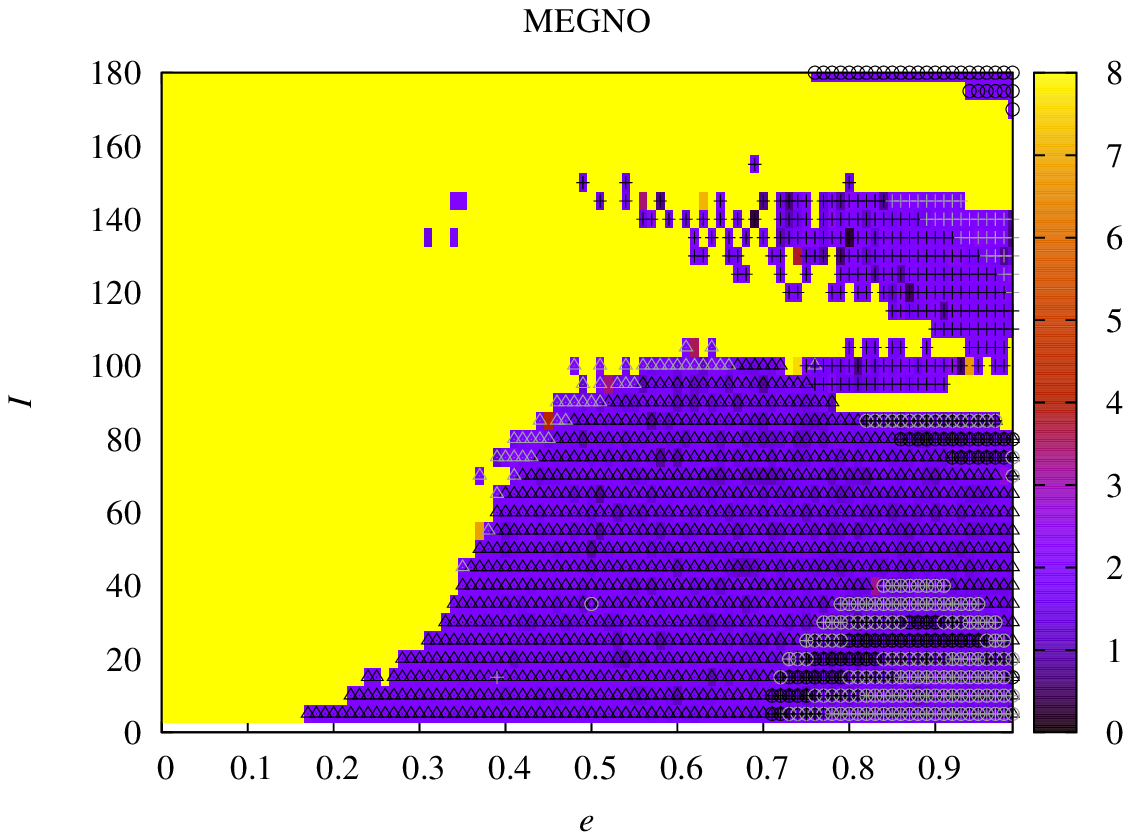}  
   \includegraphics[width=0.85\textwidth]{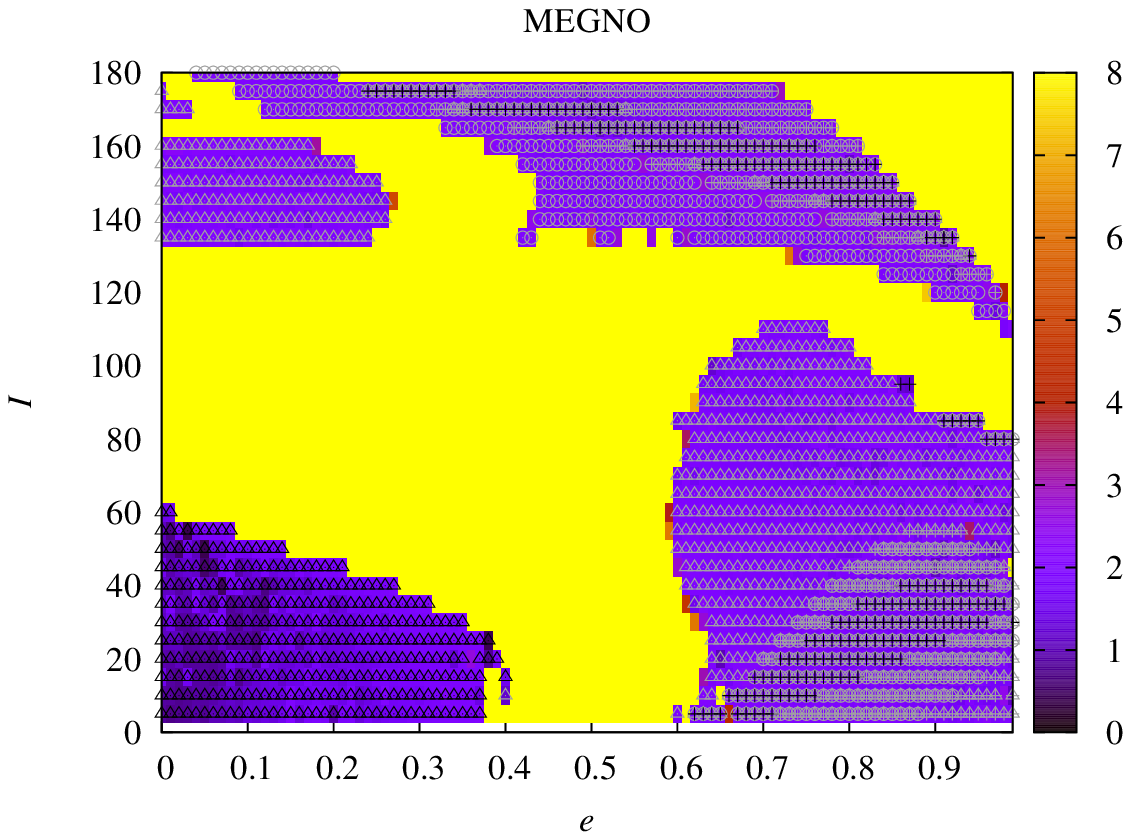}   
   \includegraphics[width=0.85\textwidth]{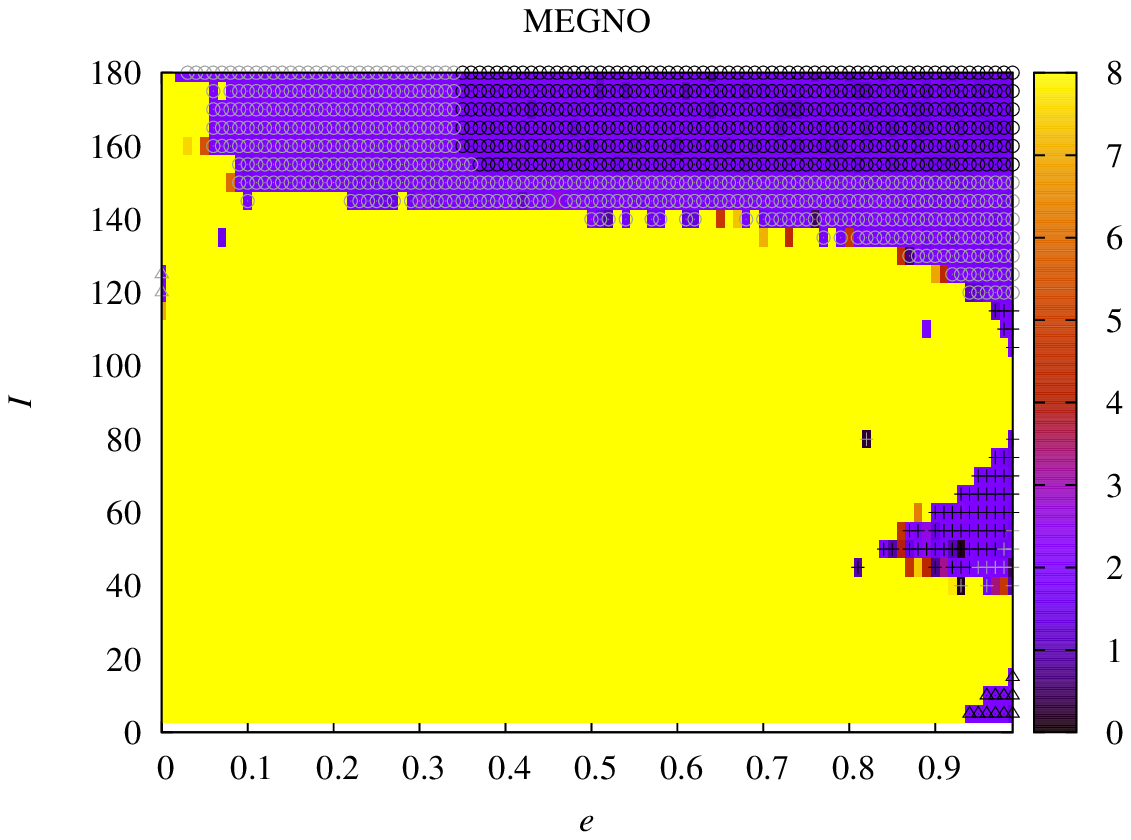}   
 \caption{Stability maps for coorbital resonance in CR3BP: initial conditions $a=1$, $\omega=90^\circ$ and $\phi=0$ (top panel), $\phi=60^\circ$ (mid panel), $\phi=180^\circ$ (low panel). The symbol $+$ marks initial conditions such that $\omega$ librates with amplitude less than $50^{\circ}$ (dark $+$) or larger than $50^{\circ}$ (light $+$). The  triangles (circles) mark initial conditions such that $\phi$ ($\phi^\star$) librate with amplitude less than $50^{\circ}$ (black) or larger than $50^{\circ}$ (gray).}    
\label{megno_w90}
\end{figure*}

In Figs.~\ref{megno_w0}\&\ref{megno_w90} we see that stable regions are mostly associated with resonant orbits such that the angles $\phi$ or $\phi^{*}$ librate, signaled respectively by triangles or circles (black symbols indicate libration amplitude less than $50^\circ$ while gray symbols indicate libration amplitude larger than $50^\circ$). 
The observed  resonant modes  are:
\begin{itemize}
\item RS: libration of $\phi$ around 0 at moderate to large eccentricity and $I\lesssim 100^\circ$;
\item T: asymmetric libration of $\phi$ around $\pm 60^\circ$ at $e\lesssim 0.4$ and $I\lesssim60^\circ$;
\item R4: libration of $\phi$ around $180^\circ$ at $e<0.3$ and $120^\circ\lesssim I<180^\circ$;
\item H: libration of $\phi$ around $180^\circ$  at $e>0.8$ and $I\lesssim 70^\circ$;
\item R1: libration of $\phi^{*}$ around 0 at wide range of eccentricities and $I>90^\circ$;
\item R2: libration of $\phi^{*}$ around $180^\circ$ at $e>0.7$ and $I\gtrsim 120^\circ$.
\end{itemize} 
Due to  output sampling some  circulating orbits are wrongly identified as high amplitude librating orbits (this explains the isolated gray circles inside the RS regions (Fig.~6\&7: top panels) and inside the T region (Fig.~6:  mid panel).
 
The  Kozai-Lidov resonance (libration of $\omega$) is identified in Figs.~\ref{megno_w0}\&\ref{megno_w90} with dark $+$ symbols (libration amplitude less than $50^\circ$) and  light $+$ symbols (libration amplitude larger than $50^\circ$) and may also ensure stability without requiring the libration of $\phi$ and $\phi^{*}$. For prograde configurations, these are the passing (P) orbits that periodically transit to the compound H-RS and RS-H modes on secular time scales \citep{Namouni1999,Namounietal1999}.

Modes RS, T and H are present in the 2D  prograde coorbital problem.  Modes R1 and R2  are present in the 2D retrograde  coorbital problem as shown in Paper I. Mode R4  only exists in the 3D coorbital problem and is the retrograde analogue to an horseshoe orbit (mode H) since $\phi$ librates around $180^\circ$.  There are regions where both $\phi$ and $\omega$ (hence $\phi^{*}$) librate which are associated with equilibria of the 3D coorbital model.  The latter orbits are identified in Figs.~6\&7 by an overlap of the triangles, circles and $+$ symbols.

When $\phi=0$ and $\omega=0$ initially (Fig.~\ref{megno_w0}: top panel) the only possible resonant orbits are: mode RS and mode R1 at moderate to large eccentricities.  The transition between RS and R1 modes does not occur exactly at $I=90^\circ$ (e.g. at $e=0.6$, RS orbits have $I\leq95^\circ$  while R1 orbits have  $I>95^\circ$).  

When $\phi=60^\circ$ and $\omega=0$ initially (Fig.~\ref{megno_w0}: mid panel) modes RS and R1 are present at moderate to large eccentricity, and mode T (tadpole orbits) occurs at $e\lesssim0.4$ and  $I\lesssim60^\circ$. Mode R4 orbits where $\phi$ librates around $180^\circ$ occur at retrograde inclination $130^\circ\lesssim I<180^\circ$ and $e<0.3$  (indeed a careful look at this panel reveals a tiny mode R4 region for nearly circular retrograde orbits slightly off the plane). 
The Kozai-Lidov resonance around $\omega=0$ occurs at the border of the region of libration in mode R4.

When $\phi=180^\circ$ and $\omega=0$ initially (Fig.~\ref{megno_w0}: low panel), mode R4 occurs at $e<0.3$ and  $I\sim 120^\circ$.
Horseshoe orbits (where $\phi$ librates around $180^\circ$) with small eccentricities and low inclinations are unstable at mass ratio $0.001$ due to the effect of close encounters with the planet (Fig.~\ref{megnoP11}: low panel). However,  such close encounters  are less disruptive when the motion is retrograde (as these occur at higher relative velocities) which explains  the stability of mode R4 orbits. 
At large eccentricities there are stable prograde horseshoe orbits (mode H)  which avoid close encounters with the planet by being always near opposition with respect to the star. Large eccentricity retrograde orbits in mode R2 are also stable.
The Kozai-Lidov resonance occurs outside the region of libration of $\phi$ or $\phi^\star$ and thus concerns passing (P) orbits. In particular, retrograde and nearly-polar orbits in that resonance are stable. 

When $\phi=0$ and $\omega=90^\circ$ initially (Fig.~\ref{megno_w90}: top panel)  mode RS is present at $e\gtrsim 0.2$ and 
$I\lesssim100^\circ$, while mode R2 occurs at $e>0.7$ and  $I\gtrsim170^\circ$.   The Kozai-Lidov resonance is present at $I\lesssim30^\circ$ inside the region of libration in mode RS,  and also at large eccentricities and inclination $100^\circ\lesssim I\lesssim140^\circ$ . Equilibria of the 3D coorbital problem occur when there is simultaneous libration in mode RS along with the Kozai-Lidov resonance since all angles ($\phi$, $\phi^\star$ and $\omega$) are stationary.
 
When $\phi=60^\circ$ and $\omega=90^\circ$ initially (Fig.~\ref{megno_w90}: mid panel) resonant orbits include mode RS at $e>0.6$ and $I\lesssim 110^\circ$, 
mode T (tadpole orbits) at $e\lesssim0.4$ and inclination $I\lesssim60^\circ$, mode R4 at $e<0.3$ and inclination $130^\circ<I<180^\circ$, mode R1 at a wide range of eccentricities and
 $I\gtrsim110^\circ$.
The Kozai-Lidov resonance   is present inside the regions of libration in modes RS  and R1. 

When $\phi=180^\circ$ and $\omega=90^\circ$ initially (Fig.~\ref{megno_w90}: low panel) possible resonant orbits are mode H at $e\gtrsim 0.95$ and $I<20^\circ$, and  mode R1 at a wide range of eccentricities and $I>100^\circ$. The Kozai-Lidov resonance occurs at $e>0.8$  and $40^\circ\lesssim I \lesssim 80^\circ$.

\begin{figure*}
\centering
  \includegraphics[width=0.85\textwidth]{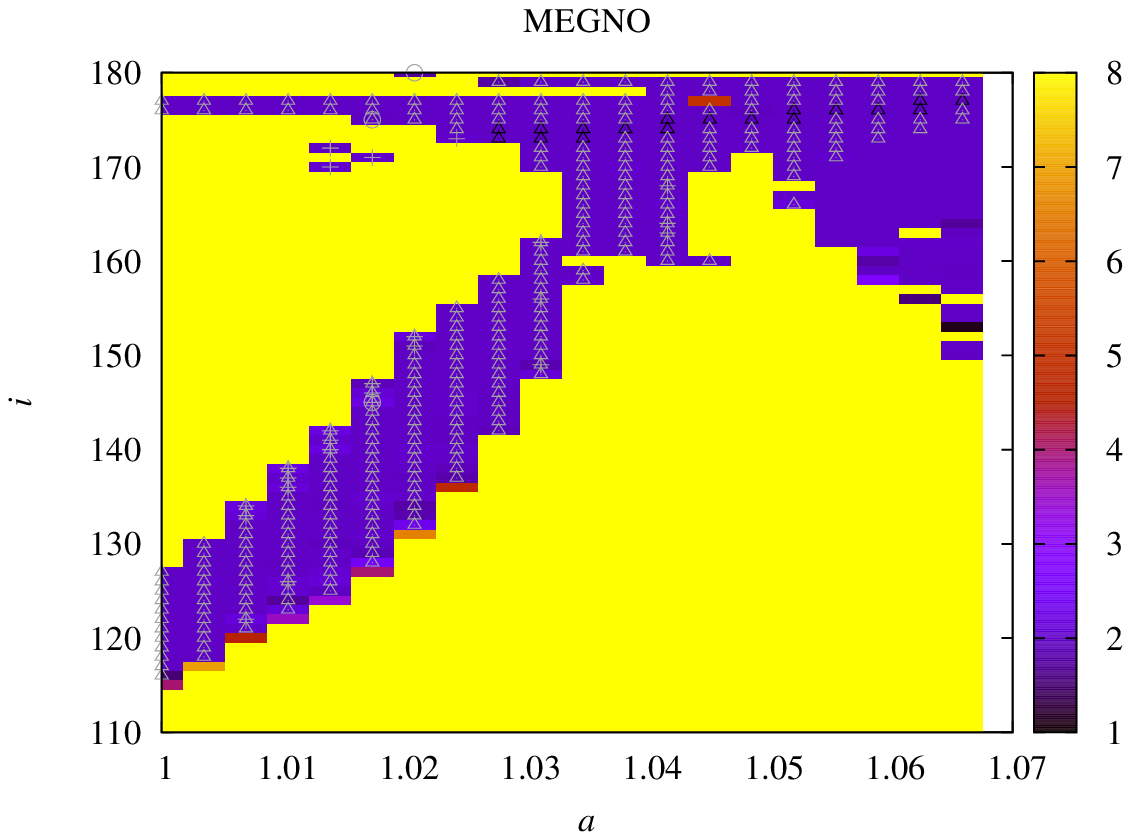}  
  \includegraphics[width=0.85\textwidth]{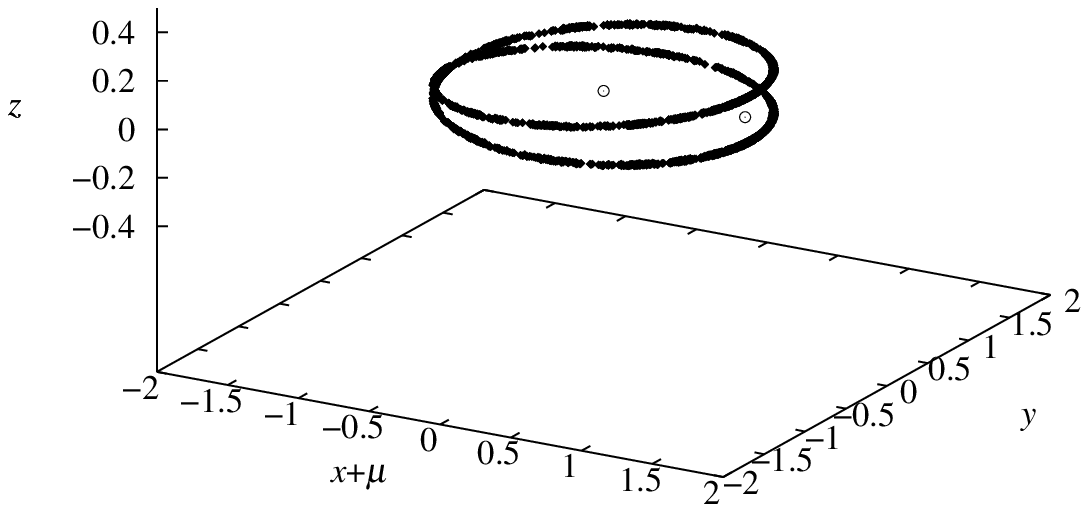}  
 \caption{Top panel: stability map in $(a,i)$ grid for initial conditions $\omega=0$, $\phi=180^\circ$ , $e=0$.  The triangles mark initial conditions such that $\phi$ librates around $180^\circ$ with amplitude less than $50^{\circ}$ (black) or larger than $50^{\circ}$ (gray). Low panel: mode R4 orbit close to exact resonance viewed in 3D synodic frame. Initial conditions: $I=175^\circ$, $\phi=180^\circ$, $\omega=0$, $a=1.04507$ and $e=0$. The star (Sun) is located at $(0,0,0)$ and the planet (Jupiter) at $(1,0,0)$. }    
\label{megno_ai}
\end{figure*}

According to Fig.~\ref{megnoR11} mode R3 is possible only at $a>1$ or $a<1$. 
In order to explore the retrograde region at small eccentricities  and to assess the stability of  mode R3 off the plane we computed stability maps with initial conditions: $\phi=180^\circ$, $\omega=0$, semi-major axis $1\le a \le 1+R_{H}$ varying at steps $0.05\,R_{H}$,  inclination $90^\circ \le I \le 180^\circ$ varying  at $1^\circ$ steps, and fixed  eccentricity: $0\le e \le 0.2$.  These initial conditions cover the regions of libration associated with mode R3 (Fig.~\ref{megnoR11}: low panel) and mode R4 (Fig.~\ref{megno_w0}: mid an low panels).  
In Fig.~\ref{megno_ai} (top panel) we show the case with initial $e=0$. We see that  mode R3 ($\phi^\star$ librating around $180^\circ$ at small eccentricities), identified by  a single gray circle located at  $a\approx 1.02$ and $I=180^\circ$, is present only in the retrograde planar problem. At inclinations slightly off the plane  this  is replaced by  mode R4  (retrograde orbit with $\phi$ librating around $180^\circ$) whose resonance centre is  identified by  black triangles in Fig.~\ref{megno_ai} (top panel).
In Fig.~\ref{megno_ai} (low panel) we show synodic frame 3D view of a mode R4 orbit corresponding to initial conditions $I=175^\circ$, $\phi=180^\circ$, $\omega=0$, $a=1.04507$ and $e=0$.
This is close to exact retrograde resonance with $\phi=180^\circ$ and corresponds to the orbit in Fig.~\ref{fig2}  approximately midway within the capture episode in mode R4. At the planet's vicinity the orbit  is either above or below the $(x,y)$ plane, thus avoiding collision. The shape in the synodic frame is closer to mode R1 ($\phi^\star=0$, as in Fig.~\ref{megnoR11}: top right panel) than mode R3  ($\phi^\star=180^\circ$, as in Fig.~\ref{megnoR11}: low right panel)

\section{Conclusion}
We investigated the mechanism of coorbital capture for retrograde orbits migrating towards a Jupiter mass planet.  This study was motivated by our previous results concerning resonance capture (Paper II) which showed that  retrograde orbits were more likely to get captured in resonance than prograde orbits. In addition, we reported  that retrograde orbits migrating inwards with small relative inclination were more likely to be captured in the coorbital resonance.  In particular, circular orbits with $I=180^\circ$  were captured into coorbital mode R3  with a $100\%$ efficiency (Paper II).  Here, we presented 2D stability maps of the retrograde coorbital resonance which show that  smooth (non chaotic)  migration capture is only possible into mode R3 for nearly circular outer orbits.

We  explained the differences between the capture mechanism and end states for  retrograde planar orbits ($I=180^\circ$ exactly) compared with retrograde orbits inclined with respect to the planet's orbital plane. While planar retrograde orbits are captured in mode R3 ($\phi^\star=180^\circ$),  for retrograde orbits only slightly off the plane capture  occurs in a 3 stage process: first mode R4 ($\phi=180^\circ$), an inclination-type resonance that does not occur in the 2D case, then the Kozai-Lidov resonance, and finally mode R1 ($\phi^\star=0^\circ$), an eccentricity-type resonance.  Our large scale ($\sim 600\, 000$) simulations from Paper II showed that 3D capture into other resonances also follows a 3-stage process.

We computed stability maps in order to identify the possible stable coorbital modes for arbitrary inclination. Such maps indicate that retrograde modes R1 and R2 persist in the three-dimensional problem while mode R3 seems to only be possible in 2 dimensions, the case studied in Paper I. This explains why capture in mode R3 for initially circular orbits occurs only when $I=180^\circ$ exactly.  An additional coorbital mode, R4, a retrograde analogue of horseshoe-type orbits, appears at low eccentricity and retrograde inclination $I<180^\circ$.   As expected, the Kozai-Lidov resonance is present in the 3D coorbital problem.  Moreover, the simulations of retrograde coorbital capture described here and in Paper II show that the Kozai-Lidov mechanism is key to the increases in eccentricity necessary for final capture into mode R1.

We saw that  retrograde coorbital resonance capture in 3D is essentially different from the 2D case even for tiny relative inclinations.  This may be a peculiarity of the retrograde coorbital resonance (we leave the detailed investigation of other resonances to a future work). Moreover, we showed that stable coorbital modes exist at all inclinations, including retrograde and polar obits. Real examples of these orbits have not yet been found. Our results imply that  high inclination coorbital companions of the solar system planets are a viable dynamical possibility and encourage future searches for such objects.

\section*{Acknowledgments}
We thank Nelson Callegari Jr for assistance with computational resources. 

\bibliographystyle{spbasic}

\bibliography{coorbitalcapture4revised}

\end{document}